\documentclass[prl,aps,twocolumn]{revtex4}
\usepackage{epsf}
\usepackage{epsfig}
\usepackage{graphics}
\begin{document}

\title{
Thermally  Fluctuating Inhomogeneous Superfluid State \\ 
of Strongly Interacting Fermions in an Optical Lattice
}

\author{Viveka Nand Singh, Sanjoy Datta  and Pinaki Majumdar}

\affiliation{Harish-Chandra  Research Institute,
 Chhatnag Road, Jhusi, Allahabad 211019, India}

\date{25 April 2011}

\begin{abstract}
The presence of attractive interaction between fermions can lead to pairing
and superfluidity in an optical lattice. The temperature needed to observe 
superfluidity is about a tenth of the tunneling energy in the optical lattice, 
and currently beyond experimental reach. However, at strong coupling the 
precursors to global superfluidity should be visible at achievable temperatures, 
in terms of fluctuating domains with strong pairing correlations.  We 
explore this regime of the attractive two dimensional fermion 
Hubbard model, in the presence of a confining potential, using a new Monte 
Carlo technique. We capture the low temperature inhomogeneous superfluid state 
with its unusual spectral signatures but mainly focus on the experimentally 
accessible intermediate temperature state. In this regime, and for the trap 
center density we consider, there is a large pairing amplitude at the center, 
spatially correlated into domains extending over several lattice spacings. We 
map out the thermal evolution of the local density, the double occupancy, 
the pairing correlations, and the momentum distribution function across this 
phase fluctuation window.
\end{abstract}

\maketitle

The development of optical lattice methods for ultracold 
atoms has opened a new vista in the study of correlated systems
\cite{jaksch1,essl-rev,bloch-rev,jaksch-hubb}, allowing
clean controllable realisations 
of strongly interacting quantum lattice models. 
Experimental achievements 
include the realisation of the superfluid (SF) to Mott insulator 
transition \cite{greiner}
in the Bose Hubbard model, and, for repulsive fermions,
the observation of Fermi surface \cite{kohl}, and the Mott insulating
phase \cite{schneider,jordens}.
For attractive interactions, there has been the evidence of
superfluidity \cite{chin}, a possible 
FFLO
state \cite{fflo}, and anomalous expansion of  the Fermi 
gas \cite{hackermuller}. 
The realisation of an antiferromagnetic state~\cite{gorelik,chiesa} 
in the repulsive
Hubbard model and of superfluidity in the attractive 
model \cite{hofs} is still awaited. The problem is with the 
achievable temperature~\cite{paiva}.

Techniques available to date can reduce the entropy,
$S$,  per particle of a Fermi gas to $\sim$ log$_e2 \approx 0.7$.  
The associated temperature is 
${\cal O}(t)$, where $t$ is the tunneling energy between
neighbouring wells in the periodic potential. 
The observation of 
superfluidity in the attractive Hubbard model 
will require cooling to
$k_BT/t \sim 0.1$. The corresponding  entropy is 
$S \sim 0.1$ \cite{paiva},
almost an order of magnitude below what is currently achievable.
What signatures would one expect of attractive interactions 
at accessible temperatures?
At weak coupling the state above $T_c$ is a normal Fermi liquid
but at strong coupling a large pairing amplitude survives to
$T \gg  T_c$ and entropy levels $S \sim$ log$_e2$. This is 
an {\it inhomogeneous, thermally fluctuating, short range correlated
state}, with striking measurable properties.

In this paper we solve the attractive (`negative $U$') Hubbard
model on large two dimensional lattices in the presence of
a confining potential. Our main results are
the following:
(i)~We access the superfluid ground state, the thermal transition,
and a wide temperature window over which the Fermi system has
large pairing amplitude but no global phase coherence.
(ii)~We illustrate how fluctuating filamentary 
`superfluid' regions survive
far above $T_c$, to temperatures $k_BT/t \sim {\cal O}(1)$,
and leave
signatures on the spatial patterns and spectral features.
We demonstrate these using a new method that handles the 
strong coupling non-perturbatively and treats the spatial 
inhomogeneity and strong thermal fluctuations exactly.

We study the attractive Hubbard model in the
presence of a harmonic potential $V_i$ in two
dimensions:
\begin{equation}
H = - t\sum_{\langle ij \rangle \sigma} c_{i \sigma}^{\dagger} c_{j \sigma} 
+ \sum_{i \sigma} (V_i - \mu) n_{i \sigma} 
- U \sum_{i} n_{i \uparrow} n_{i \downarrow}
\end{equation}
The first term denotes the nearest neighbour tunneling amplitude
of fermionic atoms on the optical lattice, the confining potential
has form $V_i = V_0(x_i^2 + y_i^2)$, $\mu$ is the chemical potential,
and $U > 0$ is the strength of attractive on-site interaction.
$x_i$ and $y_i$ are measured in units of lattice spacing $a_0$.

The spatial variation in mean value, and the
thermal fluctuation about the mean, of the 
amplitude and phase of the order parameter  
are crucial
in describing the physics of this system. Unbiased calculations
in the homogeneous limit employ determinantal quantum Monte Carlo
\cite{paiva,scalettar,moreo}
(DQMC) to access finite temperature properties, but are typically
limited to $ 10 \times 10$ lattices. That is
inadequate to clarify the interplay of correlation effects and
inhomogeneity.

We use a strategy used earlier on moderately sized systems
\cite{dubi,dag}, augmented now by a `traveling cluster' (TCA) 
\cite{tca} Monte Carlo technique that
readily allows access to system size $\sim 32 \times 32$.
We first decouple the Hubbard term in the pairing channel by
using the Hubbard-Stratonovich (HS) transformation.
We use the static  HS (sHS) approximation \cite{dubi}, 
{\it i.e},
retain spatial fluctuations of HS fields 
but ignore the time dependence.
This leads to the following effective Hamiltonian:
\begin{equation}
H_{eff}  =  H_0  +
\sum_{i} ( \Delta_{i} c_{i \uparrow}^{ \dagger} c_{i \downarrow}^{ \dagger } +
\Delta_{i}^{ \star } c_{i \downarrow } c_{ i \uparrow }) +
\sum_{i} \frac{\vert \Delta_{i} \vert^{2}}{U}
\end{equation}
where $H_0 = 
 - t\sum_{\langle ij \rangle \sigma} c_{i \sigma}^{\dagger} c_{j \sigma} 
+ \sum_{i \sigma} (V_i - \mu) n_{i \sigma} 
$ and $\Delta_i = \vert \Delta_i \vert e^{i \theta_i}$ is a complex scalar 
{\it classical} field. This model allows fluctuations in both
the amplitude and phase of the pairing field $\Delta_i$, 
and the fermions propagate typically in an  inhomogeneous background
defined by $\Delta_i$. 

To obtain the ground state, and in
general configurations $\{ \vert \Delta_i \vert, \theta_i\}$ that
follow the distribution $P\{ \vert \Delta_i \vert, \theta_i\}
\propto Tr_{c,c^{\dagger}}e^{-\beta H_{eff}}$, we use the Metropolis algorithm to
update the $\vert \Delta \vert$ and $\theta$ variables.
This involves solution of the Bogoliubov-de~Gennes (BdG)
equation \cite{bdg} 
for each attempted update. For equilibriation 
we use the  traveling cluster
algorithm \cite{tca}, diagonalising the BdG equation on a $8 \times 8$
cluster around the update site.
Global properties  like pairing field correlation, 
quasiparticle density of
states, {\it etc}, are computed via solution of the BdG equation
on the {\it full system}
for equilibrium configurations.

We explored the system at $U/t=2,~6,~12$ and the maximum (system corner)
potential $V_c \sim V_0*2*(L/2)^2 = U/2,~U,~2U$,
where the system size is $L \times L$.
This enables us to systematically study the evolution 
from weak to strong coupling, as well as weak to strong
confinement.  We focus on the strong coupling, strong
inhomogeneity case, $U=12,~V_c=24$ in this paper and will 
discuss the larger parameter set separately \cite{opt-long}.
 
In the absence of the confining potential the model is known
\cite{scalettar,moreo}  to have
a superfluid ground state for all densities $n \neq 1$, while at
$n=1$ there is the coexistence of superfluid and density wave (DW) 
correlations. The SF ground state away from $n=1$ evolves
\cite{noz} from a
Bardeen-Cooper-Schrieffer (BCS) state at $U/t \ll 1$ to a 
Bose-Einstein condensed (BEC) state of `molecular pairs' at $U/t \gg1$.  
The ground state can be reasonably accessed within mean field theory 
(MFT)  but the finite temperature predictions of MFT becomes 
increasingly inaccurate
with increase in $U$ \cite{micnas}. 
This is due to the separation of scales between
`pair formation' temperature, $T_f \sim {\cal O}(U)$, 
and pair condensation, {\it i.e},
superfluidity, which is $T_c \sim t^2/U$.
MFT captures $T_f$ but wrongly identifies it with the
superfluid transition. 

For $U/t \lesssim 1$, the state at $T > T_c$ is an uninteresting 
weakly correlated Fermi liquid. As $U/t$ increases, $T_c$ 
(in two dimensions) peaks
at $U/t \approx 5$, while $T_f$ continues to grow. A wide
`non-Fermi liquid' window opens up between $T_c$ and $T_f$, and
the system behaves like a (hardcore) Bose liquid \cite{micnas}
 for low
temperature and $U/t \gg 1$. 
In this $U \gg t$ regime, increasing $T$ leads to gradual dissociation
of the `bosons', and  paired and unpaired fermions exist in equilibrium.

The confining potential promotes an 
inhomogeneous density profile \cite{dao}, with a peak at the trap
center. If the average density near trap center is $n =1$, 
it can lead to a 
local DW pattern \cite{koga,machida}
while the SF would show up away from
the center where $n < 1$.
If the total particle number is sufficiently small so that even the
central density is $< 1$, the entire system is an
inhomogeneous SF at low temperature.
In our strong confinement problem we
 will focus on the case where the total particle number $N \approx 150$.
This leads to a trap center density $n=0.9$, optimising the $T_c$
for our choice of $U$ and $V_c$.
For weaker confinement a larger $N$ can be used.

\begin{figure}[t]
\centerline{
\includegraphics[width=6.5cm,height=4.0cm,angle=0]{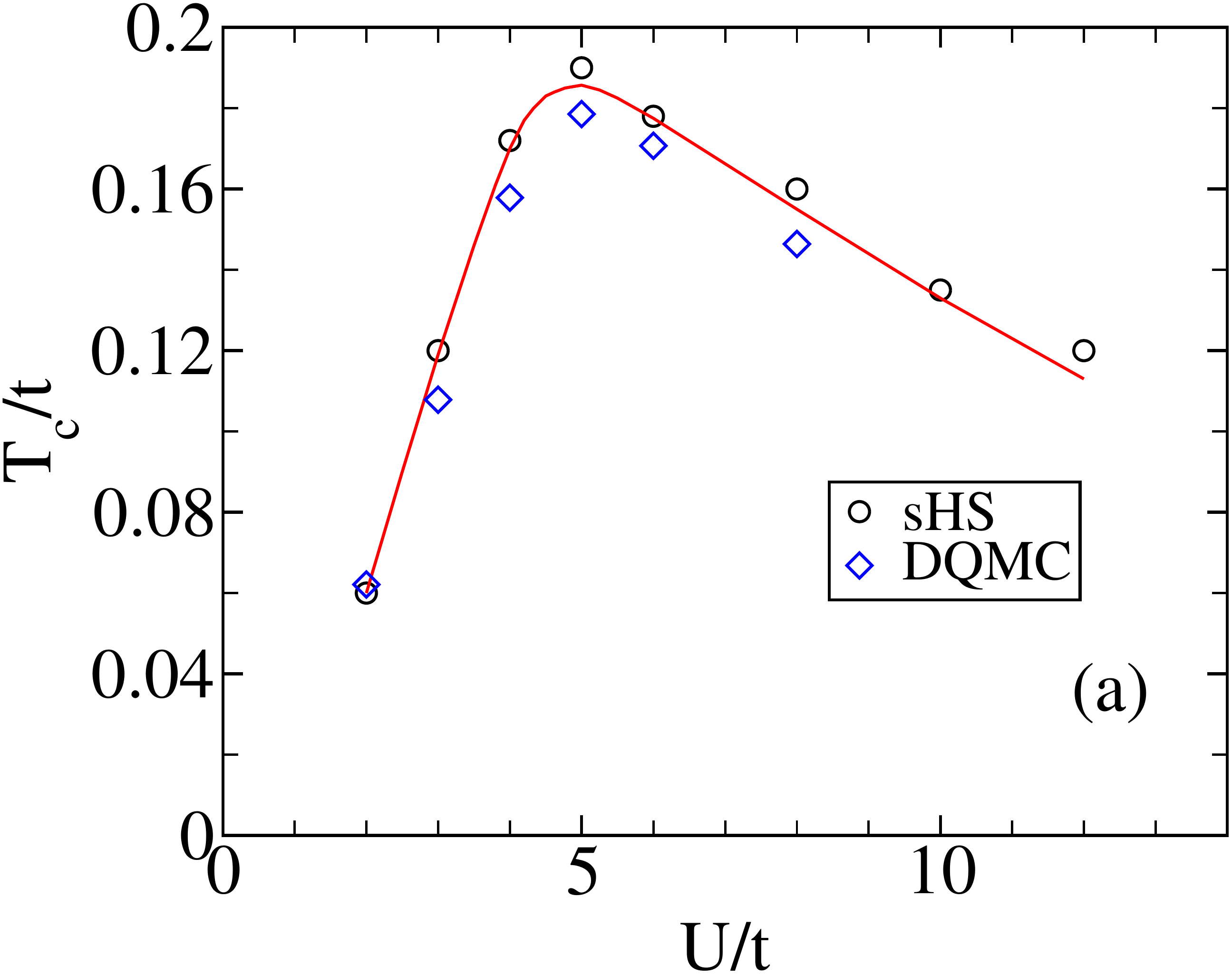} }
\centerline{
\includegraphics[width=6.3cm,height=4.0cm,angle=0]{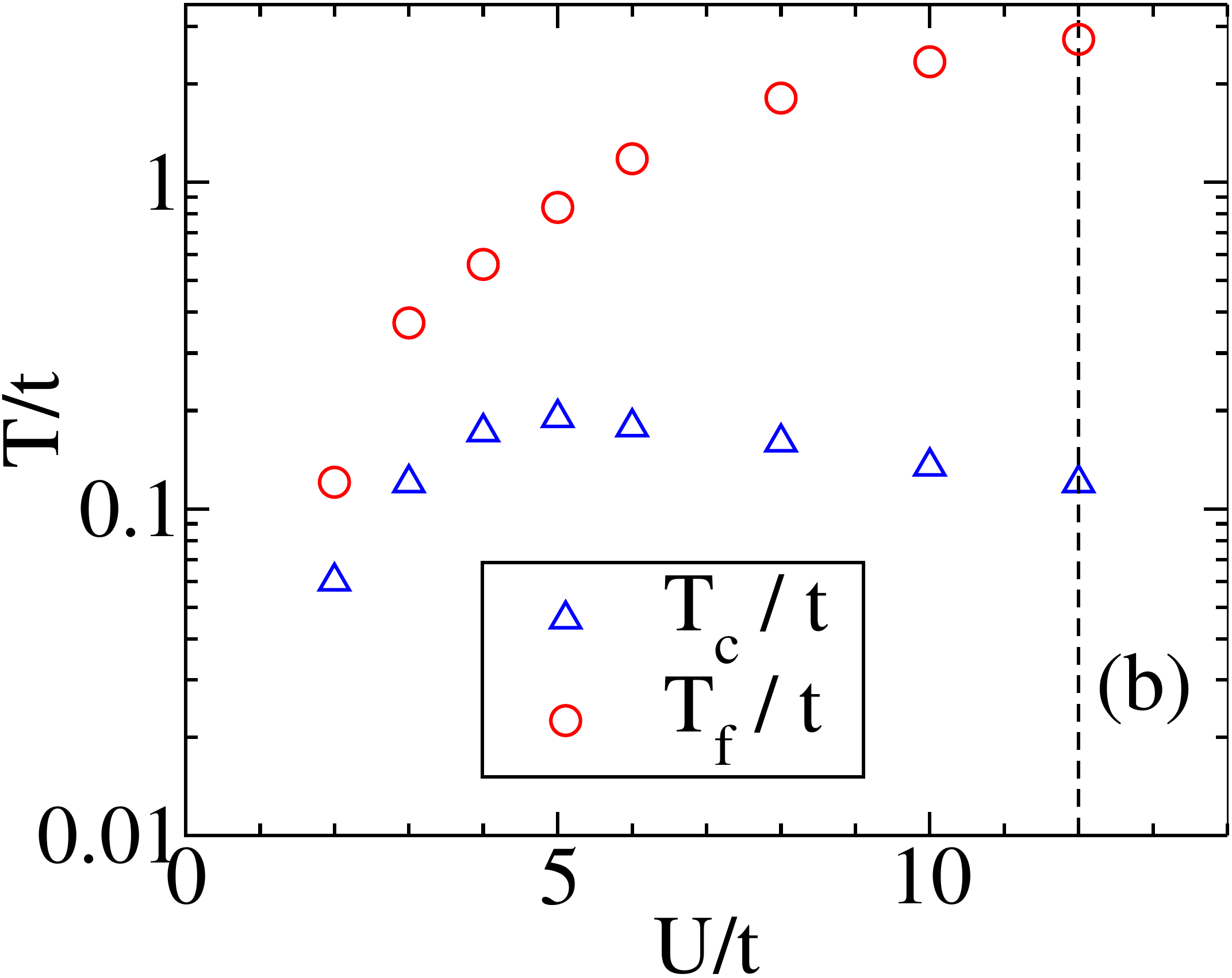} }
\centerline{
\includegraphics[width=6.5cm,height=4.0cm,angle=0]{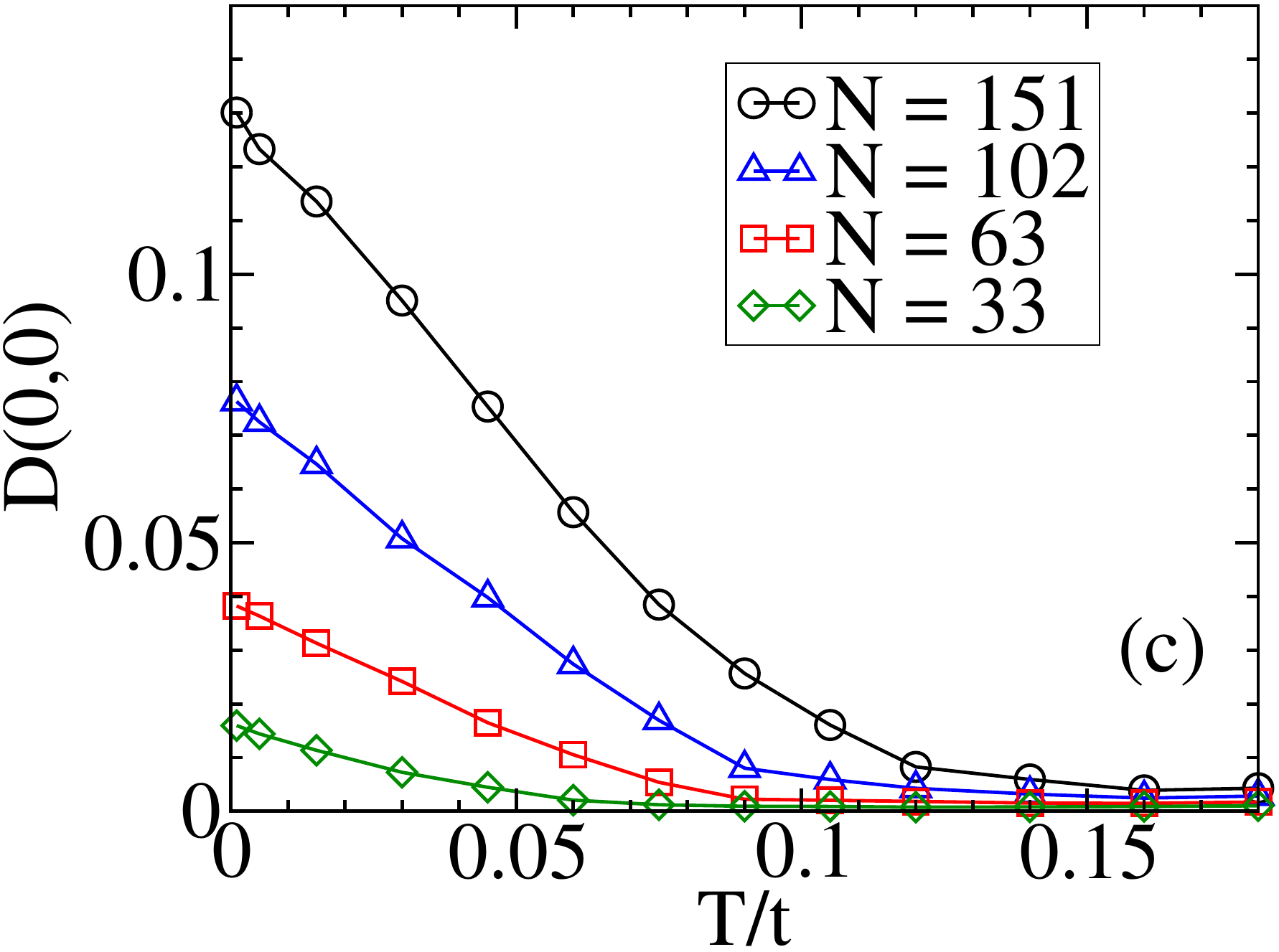} }
\caption{Colour online: (a).~Comparison of the $T_c(U)$ at $n=0.7$
on `flat' $10 \times 10$ lattices, obtained by two different
techniques, the DQMC and our static HS approximation.
(b).~Schematic phase diagram in the flat case,
indicating the window between
the SF transition ($T_c$) and  `pair formation' scale
($T_f$) that grows with increasing $U$. The $y$ scale
is logarithmic to include the widely different scales of $T_c$ and
$T_f$. The vertical dotted line indicates the $T$ dependence
that we explore.
(c).~The growth of superfluid correlations at $U/t=12$,
$V_c/t=24$, as
indicated by the zero momentum component of the pairing field
correlation. Both the onset temperature and $T=0$ value are
suppressed at lower $N$.
System size $32 \times 32$.
}
\end{figure}

Fig.1.(a) compares the result of full DQMC calculation \cite{paiva}
with that of the sHS scheme implemented via TCA. 
The comparison on $10 \times 10$ lattices, with $V_0=0$  and $n \sim 0.7$,
shows that our method
captures the overall trend in $T_c(U)$ and is even quantitatively
accurate. In our understanding the agreement is due to the
inclusion of the key thermal phase fluctuations within the static
HS theory.  This gives us confidence in the method when 
applied to large lattices
and the presence of a potential.

Fig.1.(b) highlights the two key scales from
the $V_0=0$
problem, at $n=0.7$, that are relevant for us:
(i)~the non-monotonic $T_c$ scale whose maximum is roughly $0.18t$
and, (ii)~the pair formation temperature $T_f$ as defined below.
We know that the pair binding energy is ${\cal O}(U)$ when
$U/t$ is large. To fix the prefactor, we 
{\it define} $T_f = 2\Delta_g(0)/3.5$, where $2\Delta_g(0)$ is the
$T=0$ gap in the quasiparticle spectrum. 
For $U/t \ll 1$, both superfluidity and the pairing amplitude vanish
when $T = 2\Delta_g(0)/3.5$ and $T_c = T_f$ by definition.
At strong coupling $2\Delta_g \sim U$, so $T_f \propto U$.

Fig.1.(c) shows the growth in the ${\bf Q} =\{0,0\}$ component of the
pairing field correlations,
$D({\bf Q}) = \sum_{ij} \vert \Delta_i \vert \vert \Delta_j \vert
 cos(\theta_i - \theta_j) 
e^{i {\bf Q}.({\bf r}_i - {\bf r}_j)} $. This is  non zero when 
the amplitude $\vert \Delta_i \vert $
 is finite over some region and the phases
$\theta_i$ are correlated. This in turn promotes a non zero value of 
$ \langle \langle 
c^{\dagger}_{i\uparrow} c^{\dagger}_{i \downarrow}
\rangle \rangle$ and a finite value for $\chi({ij}, T)  =  
\langle \langle 
c^{\dagger}_{i\uparrow} c^{\dagger}_{i \downarrow}
c_{j\downarrow} c_{j \uparrow}
\rangle \rangle$.  
We therefore use $D(\{ 0,0\}, T)$ 
as indicator of the SF transition.
This is shown for choices of particle number $N$ 
that lead to trap center density $n_i \le 1$.
The $T_c$ scale, at $N \sim 150 $ is roughly $0.12t$. At lower $N$
the onset temperatures are lower and the strength of pairing field
correlation at $T=0$ is also smaller.

Accessible temperatures are still $\gg T_c^{\max} \sim 0.18t$ so
one would have to look for non trivial interaction effects at $T > T_c$.
At weak coupling 
the $T > T_c$ state is uninteresting. However, for $U/t \gtrsim 5$,
the window between $T_c$ and $T_f$ is wide and well accessible with
present cooling 
techniques. We highlight the particularly wide window at
$U=12$ that reaches from $T/t \sim 0.1-3$.

\begin{figure}[t]
\centerline{
\includegraphics[width=2.5cm,height=2.5cm,angle=0]{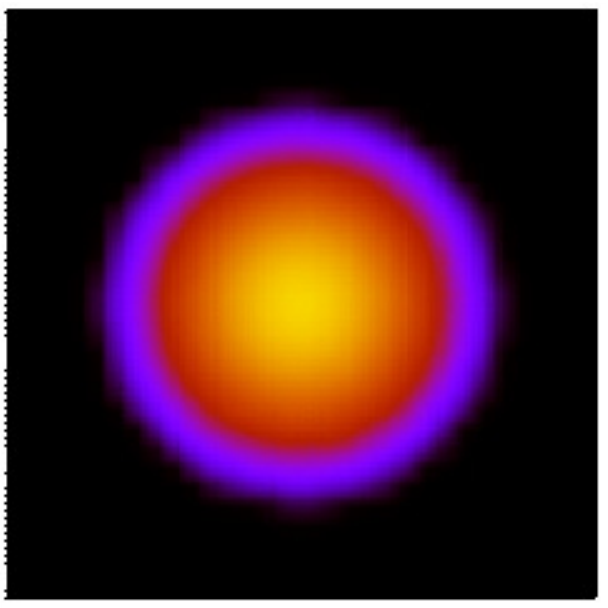}
\includegraphics[width=2.5cm,height=2.5cm,angle=0]{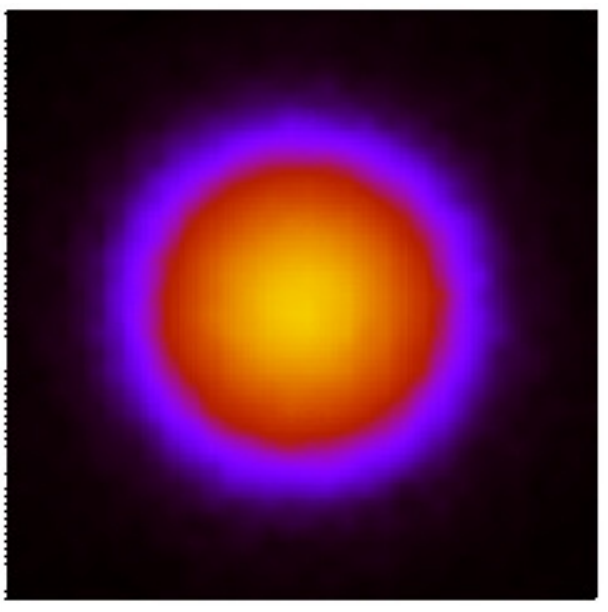}
\includegraphics[width=2.5cm,height=2.5cm,angle=0]{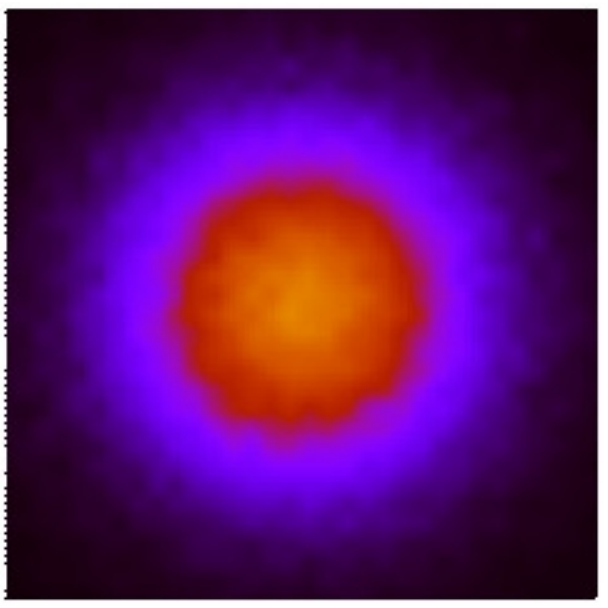}
\includegraphics[width=.6cm,height=2.5cm,angle=0]{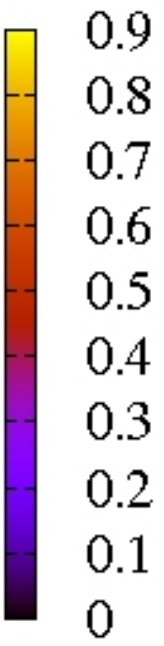}
}
\centerline{
\includegraphics[width=2.5cm,height=2.5cm,angle=0]{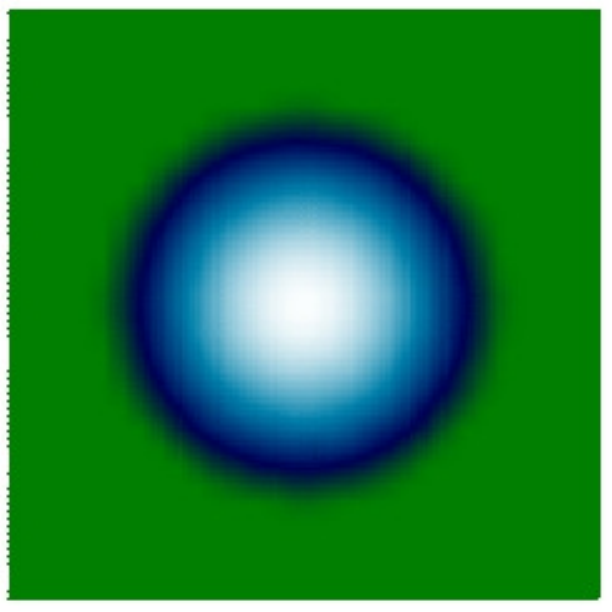}
\includegraphics[width=2.5cm,height=2.5cm,angle=0]{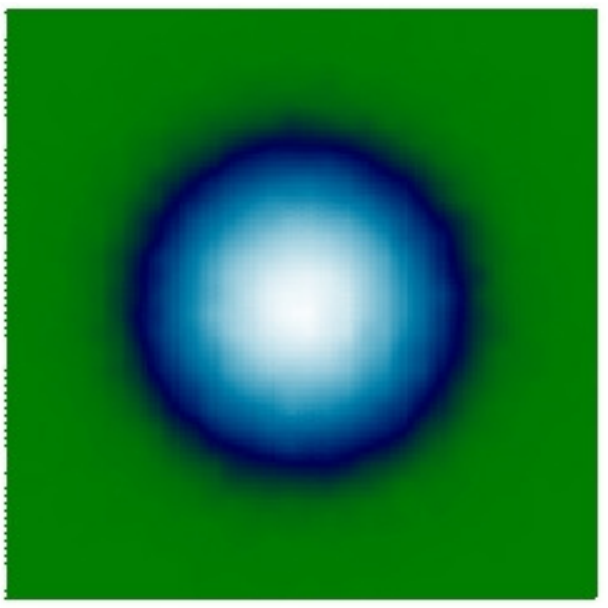}
\includegraphics[width=2.5cm,height=2.5cm,angle=0]{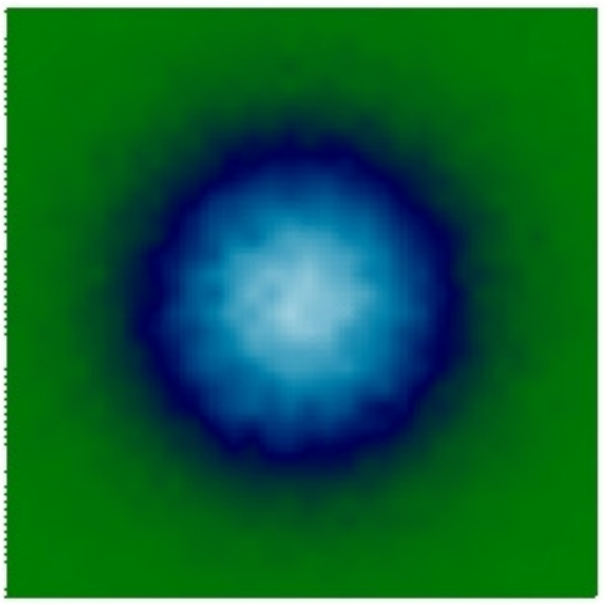}
\includegraphics[width=.6cm,height=2.5cm,angle=0]{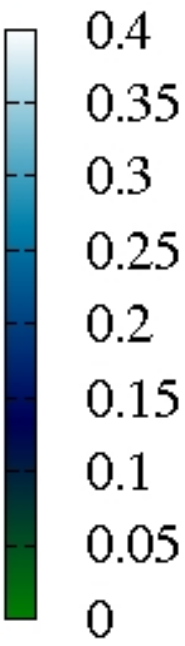}
}
\centerline{
\includegraphics[width=2.5cm,height=2.5cm,angle=0]{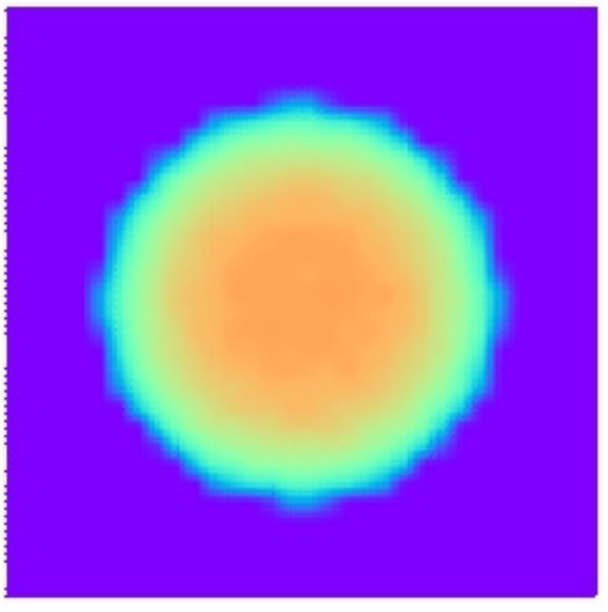}
\includegraphics[width=2.5cm,height=2.5cm,angle=0]{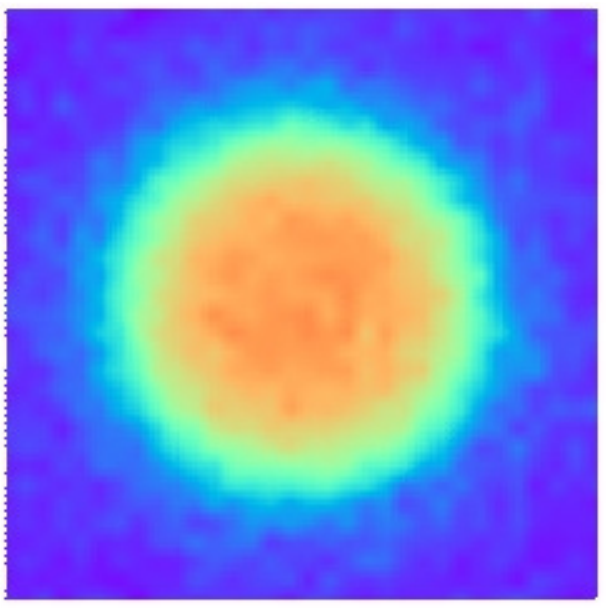}
\includegraphics[width=2.5cm,height=2.5cm,angle=0]{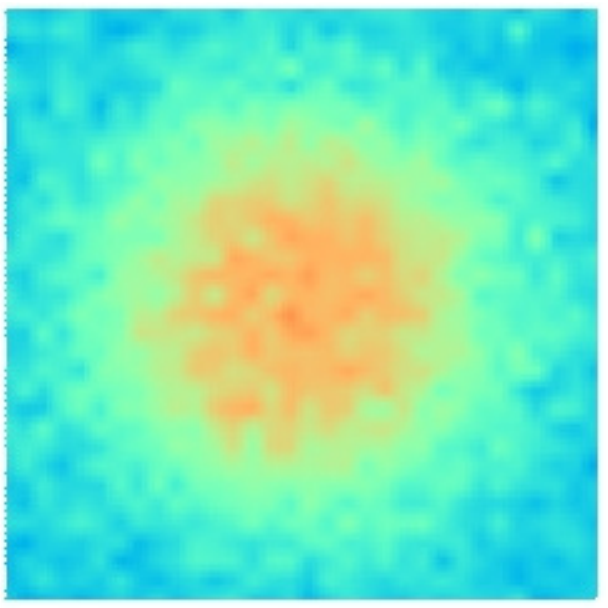}
\includegraphics[width=.6cm,height=2.5cm,angle=0]{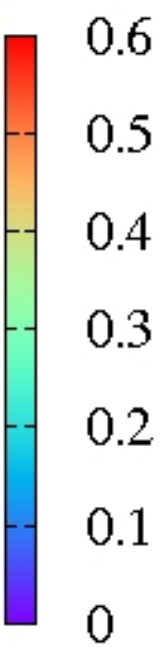}
}
\centerline{
\includegraphics[width=2.5cm,height=2.5cm,angle=0]{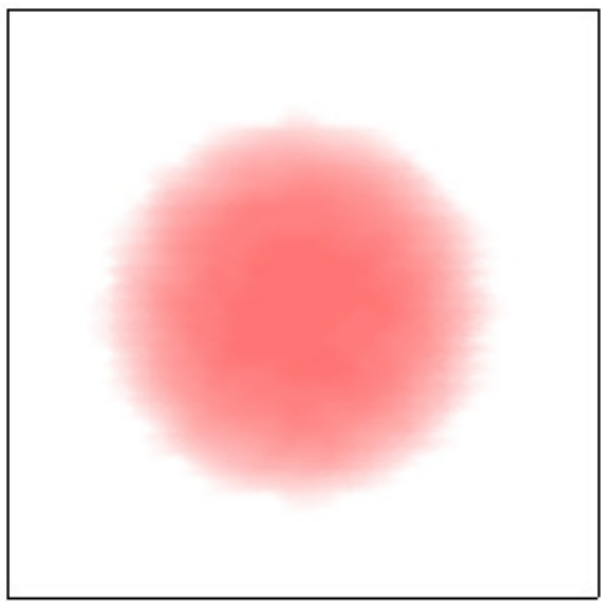}
\includegraphics[width=2.5cm,height=2.5cm,angle=0]{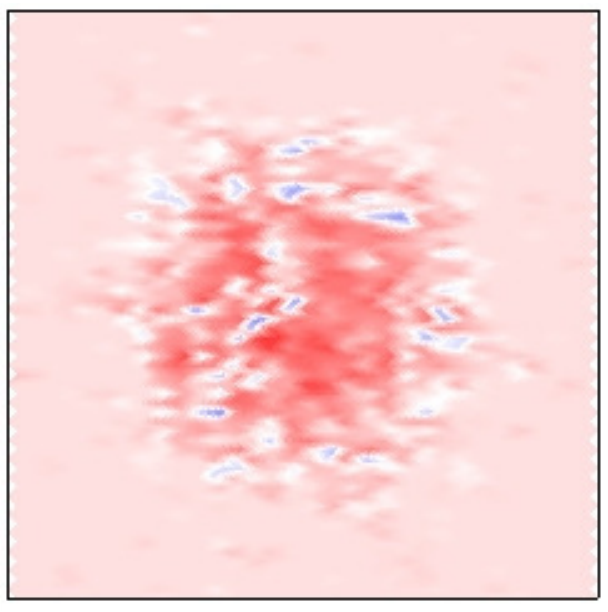}
\includegraphics[width=2.5cm,height=2.5cm,angle=0]{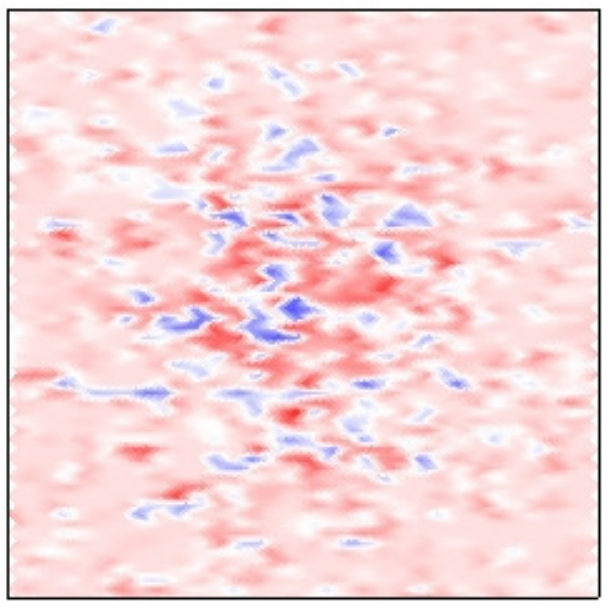}
\includegraphics[width=.6cm,height=2.5cm,angle=0]{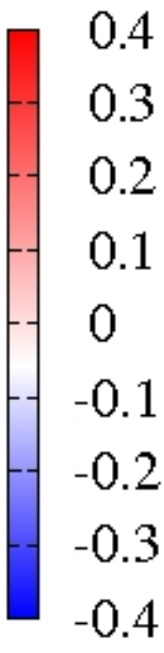}
}
\caption{Colour online: 
Temperature dependence of spatial patterns.
The results are all for $N \approx 150$. The first column
is for $T/t=0.001$ (the ground state), the second column for $T/t =0.090$
(roughly below $T_c$) and the third column at $T/t =0.50$, deep in the
fluctuating regime. 
The first row shows $n_i$, the second shows the double occupancy $d_i$,
the third shows the pairing field magnitude $\vert \Delta_i \vert$,
 the fourth shows
the nearest neighbour pairing field correlation.
System size $32 \times 32$.
}
\end{figure}

We chose $\mu$ such that the maximum density, which occurs at the
trap center, was always less than $1$, and the ground state of
the system does not involve any DW order. Our ground state is
always an inhomogeneous SF. Fig.2, left column,
shows the spatial
patterns in the `ground state' ($T=0.001t$).
For our choice of $\mu$, the density at the center is
$\sim 0.9$. 
The double occupancy
$d_i = \langle \langle n_{i \uparrow} n_{i \downarrow}
\rangle \rangle$ follows a similar profile and is almost
{\it double} the noninteracting value $(n_i/2)^2$ due to the
strong interaction. The pairing field amplitude is also
largest at the center (in the uniform case the
pairing amplitude increases 
with $n$ from $n=0$ to $n \lesssim 1$). The phase correlations are
near perfect in the ground state, as the bottom row, left column
indicates.  

The central and right column in Fig.2  highlights
the thermal evolution,
with the results averaged over 40 configurations.
The middle column is for $T/t=0.09$, and the right for $T/t=0.50$.
The global order parameter for superfluidity vanishes at $T/t \sim 0.12$
but, as expected at large $U/t$, the pairing amplitude still survives.
From $T=0.001t$ (left column) 
to $T \sim T_c$ (middle column)  there is no
significant change in the density pattern, the double occupancy,
or the pairing amplitude.
The pairing correlation (bottom row) is still
dominantly positive at $T/t=0.09$ but with 
hints of small (minority) domains.

At the highest temperature, $T=0.50t$, right column, 
where we expect the average entropy
to be $\sim 0.5$ per particle, the $n_i$ pattern is significantly
broader and the associated $d_i$ is more diffuse (with a slightly
lower trap center value). The pairing amplitude is still significant,
although the averaging has not completely restored the circular
symmetry. The pairing correlation reveals a strong short range feature,
and a filamentary pattern with lengthscale $\sim 5a_0$. The 
correlations are stronger, overall, near the central part, 
but now have some strength towards the periphery also due to 
the density broadening.  

\begin{figure}[t]
\centerline{
\includegraphics[width=6.2cm,height=7.0cm,angle=0]{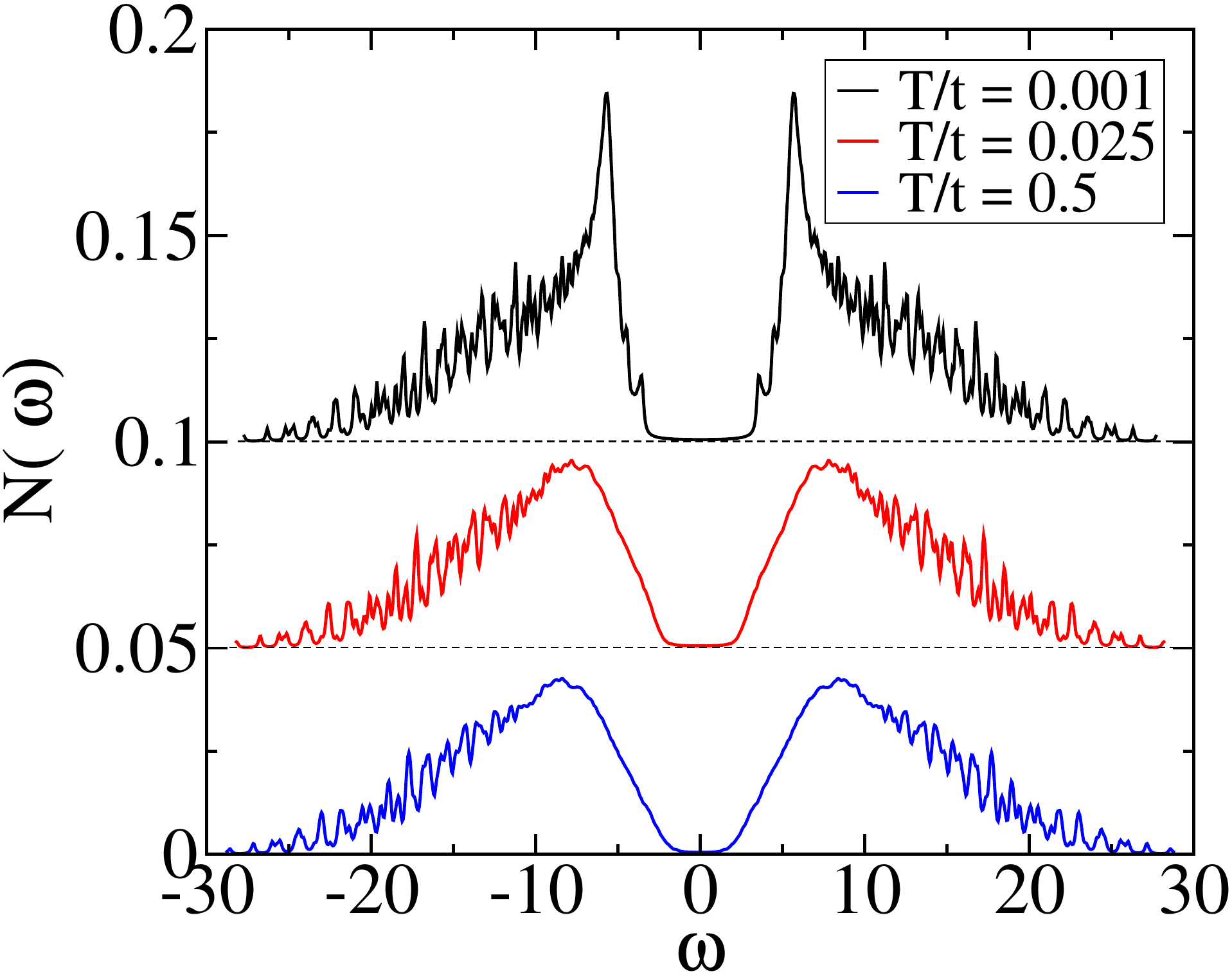}
}
\caption{Colour online: Quasiparticle density of states  for
$T/t=0.001,~0.25,~0.50$. $T_c \approx 0.12$, so the results are
approximately for $T =0,~2.5T_c$ and $5T_c$. The gap and the spiky features
survive over this temperature window and are related, respectively, to
`preformed pairs' and the level quantisation due to the confining
potential. System size $32 \times 32$.}
\end{figure}
A direct measure of the correlated phase is the quasiparticle density
of states (DOS),
$N(\omega) = \langle \langle \sum_n \delta(\omega - E_n)
\rangle \rangle $, shown in Fig.3 for $T/t=0.001,~0.25,~0.50$.
$E_n$ are the BdG eigenvalues in the
equilibrium $\Delta_i$ background. 
At $T=0.001$ there are three
noteworthy features: (i)~the pairing gap $\approx U$, (ii)~the
coherence peaks at the gap edge, reminiscent of `flat' systems,
and (iii)~the `spiky' features that arise from the 
quantisation of energy levels in this `stiff' trap.  
We have checked that the sharp levels survive, but become more
regularly spaced, even in the non-interacting case.
It is now possible to measure the spectral function via
photoemission \cite{jin-pes} in cold atom experiments.

\begin{figure}[b]
\centerline{
\includegraphics[width=2.5cm,height=2.5cm,angle=0]{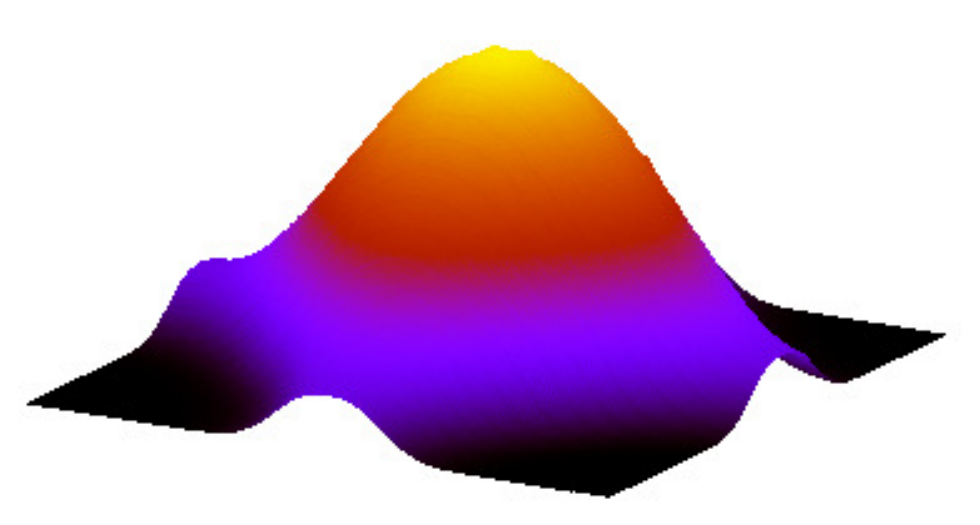}
\includegraphics[width=2.5cm,height=2.5cm,angle=0]{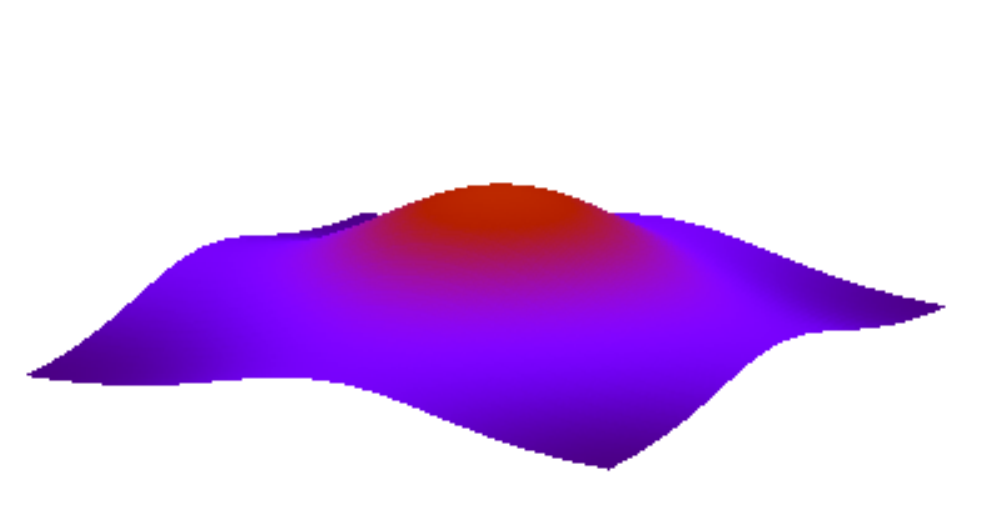}
\includegraphics[width=2.5cm,height=2.5cm,angle=0]{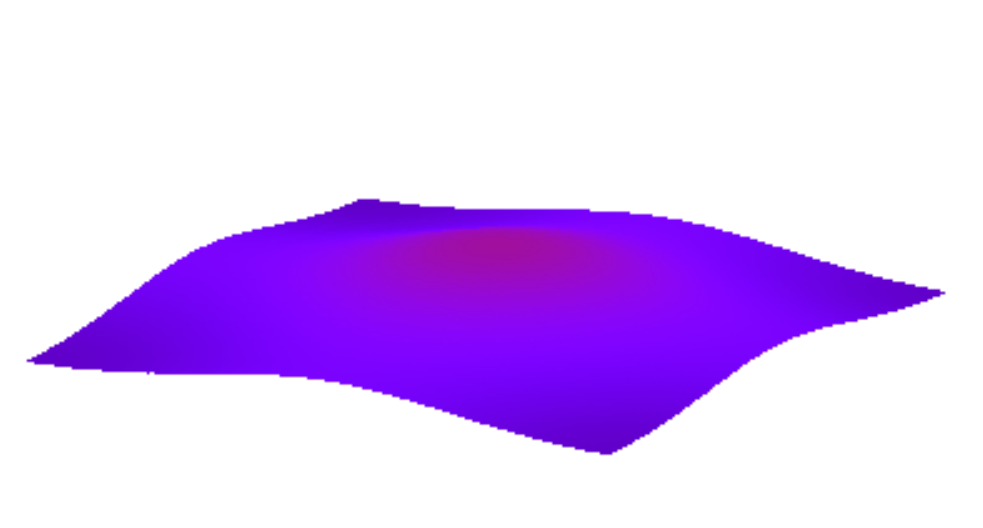}
\includegraphics[width=1.0cm,height=2.5cm,angle=0]{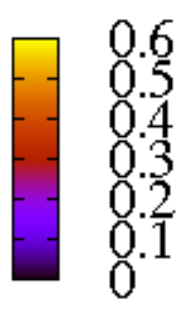}
}
\caption{Colour online: Momentum distribution function, $n(k_x,k_y)$.
The left panel shows the $n(k_x,k_y)$ for a {\it non-interacting}
fermion gas in the trap, with $N=150$ and $T=0$. Although
there is expectedly no sharp `Fermi surface', due to the background
potential, $n(k_x,k_y)$ has strong momentum dependence. Middle
panel is for trapped interacting fermions at $T/t=0.001$. Right
panel is the trapped interacting system at $T/t=0.50$}
\end{figure}
We observed that with increasing $T$ the coherence 
features get wiped out and vanish
by the time $T \sim T_c/2$. The pairing gap, however, survives
but with two modifications. The region over which $N(\omega)=0$ 
now shrinks (the gap lessens) but with the loss of the coherence
peaks there is a loss in band edge spectral weight. The quantised
features still survive but are more diffuse. 
This gapped spectrum is visible even at $T/t=0.50$, like the Mott
gap in the positive $U$ Hubbard model.

Let us discuss the momentum distribution function $n(k_x,k_y)$,
Fig.4, as the final signature of correlation physics.
This can be measured from the velocity distribution of the
gas by switching off the trap. 
The left panel in Fig.4 shows $n(k_x,k_y)$ for the ground state
of the {\it non-interacting} trapped gas (at same $N$). While
there is no Fermi surface (FS) there is a strong momentum 
dependence. $n(k_x,k_y)$ is very distinct in the 
interacting case: flat and broad with only a weak central peak
in the
ground state, and essentially flat at $T/t=0.50$. 
Tuning the Feshbach resonance across the BCS-BEC
crossover should observe this 
broadening.

{\it Conclusions:}
We have studied the attractive Hubbard model at strong coupling 
in the presence of 
a harmonic confining potential. Our non perturbative
results highlight the destruction of global superfluid order at
fairly low temperature but the survival of nanoscale fluctuating
`paired' regions to high temperature. They leave an imprint on
the spectral density and momentum distribution which 
serve as precursors to global  coherence.

{\it Acknowledgements:}
We acknowledge use of the Beowulf Cluster at HRI and thank S. Tarat for
discussions.
PM acknowledges support from a DAE-SRC Outstanding Research Investigator
Award, and the DST India~(Athena).

\end{document}